\documentclass[a4paper,12pt]{article}
\usepackage{indentfirst}
\usepackage{listings}
\usepackage{graphicx}
\usepackage{subcaption}
\usepackage{epstopdf}
\usepackage{amsmath}
\usepackage{xcolor}
\usepackage{pgfplots}
\usepackage{tikz}
\usepackage{url}
\usepackage{hyperref}
\usepackage[inner=2.5cm,outer=2.5cm,top=2.5cm, bottom=2.5cm]{geometry}
\RequirePackage{geometry}
\usepgfplotslibrary{units}
\usetikzlibrary{arrows, positioning}
\usetikzlibrary{shapes}

\definecolor{bblue}{HTML}{4F81BD}
\definecolor{rred}{HTML}{C0504D}
\definecolor{ggreen}{HTML}{9BBB59}
\definecolor{ppurple}{HTML}{9F4C7C}

\title{\textbf{Static Code Analysis with CodeChecker}}
\author{G\'{a}bor Horv\'{a}th, R\'{e}ka Nikolett Kov\'{a}cs, Rich\'{a}rd
  Szalay, \\ Zolt\'{a}n Porkol\'{a}b \\ {\normalsize E\"{o}tv\"{o}s Lor\'{a}nd
  University,} \\
  {\normalsize Dept.\,of Programming Languages and Compilers,} \\
  {\normalsize Budapest, Hungary} \\
  {\normalsize \texttt{\{xazax, rekanikolett\}@caesar.elte.hu,}} \\
  {\normalsize \texttt{szalayrichard@inf.elte.hu, gsd@elte.hu}}
  \and Gy\"{o}rgy Orb\'{a}n, D\'{a}niel Krupp \\
  {\normalsize Ericsson Hungary Ltd.} \\
  {\normalsize \texttt{\{gyorgy.orban, daniel.krupp\}@ericsson.com}}}
\date{Jan 22--26, 2018.}

\begin{document}

\clearpage\maketitle
\thispagestyle{empty}

\begin{center}
  \includegraphics[height=.4\textheight,keepaspectratio]{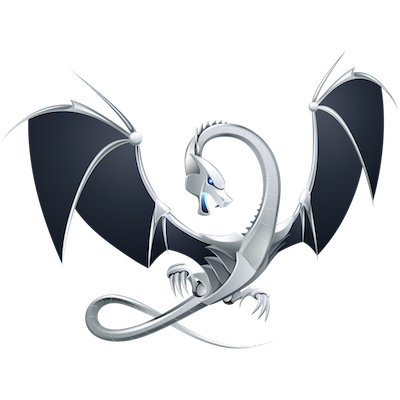}
\end{center}

\section{Introduction}

Static analysis is the analysis of a program without executing it that
is carried out by an automated tool.
It is a widely used approach for several problems. It is used for
optimising code, large-scale refactoring, code metrics, code visualisation and
several other purposes. Unfortunately, there are some strong theoretical limits
on the power of such analysis -- for example the halting problem.Notwithstanding
the limitations, it proved to be a practical tool and also continues to be an
active field of research.

Maintenance costs of a software increase with the size of the
codebase. Static analysis
has a great impact on reducing maintenance costs of complex software \cite{Zhivich2009}. For
example, compilers can detect more and more optimisation possibilities
statically and these optimisations make it possible to develop software using
high-level language features. Static analysis, however, is also a great approach
to find bugs and code smells \cite{Bessey2010}. The earlier a bug is detected
the lower the cost of the fix \cite{fixcost}. This makes static analysis a useful and
cheap supplement to testing. Moreover, some properties of the code cannot be
tested, such as compliance with coding conventions.

\emph{Symbolic execution} \cite{Hampapuram2005} is a major \emph{abstract interpretation} technique.
During symbolic execution, the program is being interpreted, without any knowledge
about the run-time environment. Symbols are used to represent unknown values
(the value might be only known at run-time, e.g. user input), and symbolic
calculations are carried out on them. The analyser attempts to enumerate all the
possible execution paths. Due to the vast number of possible paths, there are
certain heuristics regarding where to cut the analysis and how to deal with missing information and
approximate certain aspects of the program.

The open source \emph{Clang} compiler \cite{Lattner2008} has a component called \emph{Static Analyzer} (SA).
It implements a powerful symbolic execution engine for C, C++, and Objective-C.
It also implements a lot of useful checks to detect programming errors.

Clang Tidy is an open source tool to find code smells and refactor code.
This tool is used extensively within Google during code reviews. 
It is using a domain specific pattern matching language on the abstract
syntax tree.

CodeChecker is an open source project to integrate different static analysis
tools such as the Clang Static Analyzer and Clang Tidy into the build system,
CI loop, and the development workflow. It also has a powerful issue management
system to make it easier to evaluate the reports of the static analysis tools.

CodeCompass~\cite{CodeCompass} is a tool that also utilises static analysis
approaches to aid code comprehension on large-scale projects.

\section{Compilers}

It is worth to study compilers since they are everywhere. You will find multiple
compilers in the browsers, there are also some in the graphics card drivers,
your spreadsheet application, your machine learning framework, and so on.
Knowing the basics of how compilers work might help you understand the tools
that we use on a daily basis.

The most widely used tools that use static analysis are compilers. For this
reason, historically lots of the algorithms and data structures that are used in
such tools originate from compiler research. This section introduces some
of the compiler related concepts that are used throughout this handout.

The compilation process usually starts with lexical analysis (\emph{lexing}). During lexing the
compiler identifies lexical elements in the source text such as operators,
identifiers, and literals. The result of the lexing phase is a stream of tokens.
Tokens are lexical elements that consist of source locations and a kind.

\begin{figure*}
\centering
\includegraphics[width=140mm]{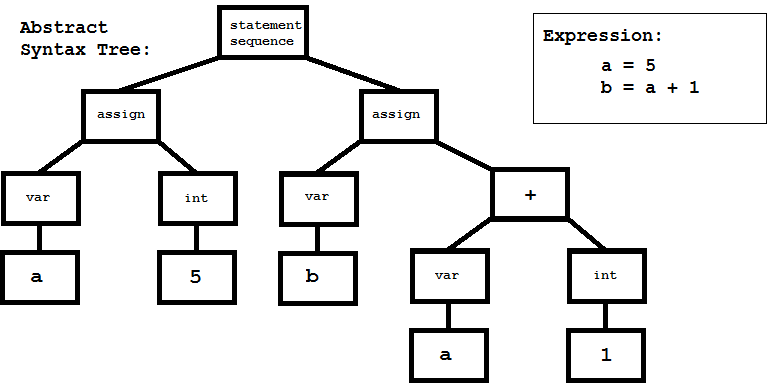}
\caption{Abstract syntax tree}
\label{fig:ast}
\end{figure*}

The token stream is parsed. During parsing the compiler builds the abstract
syntax tree (AST). This tree data structure represents the relationships between
the lexical elements. The syntax tree is called abstract because some of the
lexical elements might not appear explicitly (like parentheses) because this
information might be available implicitly in the structure of the tree. It
may also contain implicit constructs that do not appear in the source text.
On the other hand parse trees or concrete syntax trees contain all the
information about the source text (including whitespaces and comments),
so the original text could be restored from this data structure using pretty-printing.
It is common to not build a parse tree at all, only build the abstract
syntax tree instead. A parse tree can be useful when precise pretty printing support is
required. An example AST can be viewed in Figure~\ref{fig:ast}.

After parsing the source code, sometimes some syntactic constructs are reduced
to a simplified form. This method is called \emph{desugaring}. Basically, the language is
reduced to a core language which is much simpler to deal with. This core
language has the same expressive power than the original one but it is not as
convenient to write.

A process closely related to desugaring is called cannibalisation. In some
cases, there are more ways to express the same concept in a language or an
intermediate representation. For example, there might be more ways to
represent loops. Cannibalisation will try to transform different representations
to the same form so later phases dealing with these construct are less complex.

After parsing, the next phase is semantic analysis. This phase consists
of type checking and scope checking. The AST is decorated with type information
during this step. The resulting data structure contains enough information to
generate executable code.

In some languages, in order to resolve ambiguities in the grammar, it is necessary
to have some type information available. In this case, parsing and 
semantic analysis is done in the same phase. 

During the compilation, errors might be found in the source text. There are
multiple kinds of possible errors depending on which phase is responsible to
for finding them. The most common ones are lexical errors, syntax errors, and type errors.

Nowadays compilers provide the user with advanced warning messages, for example,
they can detect variables that are uninitialised on an execution path. In order
to carry out such analysis, it is often beneficial to build the control flow
graph (CFG).  The
CFG is a graph that has basic blocks as its nodes and the edges are the possible
jumps between those blocks.
A basic block is a fragment of code that contains only sequentially executed
instructions. This means that it contains no jump instructions and every such
instruction can only jump to the beginning of a basic block.

Since creating a parser for a language requires heroic effort, it is more
common to build static analysis tools on the top of an existing compiler. This
way the tool would use the internal representation of the compiler, for example,
the AST or the CFG instead of introducing its own parser.

\section{LLVM}

The LLVM Project is a collection of modular and reusable compiler and toolchain technologies.
Despite its name, LLVM has little to do with traditional virtual machines.
The name "LLVM" itself is not an acronym; it is the full name of the project.

The LLVM core libraries provide a modern source and target-independent 
optimiser and code generators for many popular architectures. 

\begin{figure*}
\centering
\includegraphics[width=160mm]{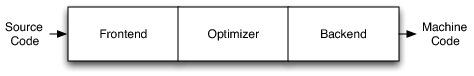}
\caption{Bug list view in CodeChecker}
\label{fig:SimpleComp}
\end{figure*}

Traditionally compilers have three parts. The front-end is parsing the source
code and creates an intermediate representation. The middle end or optimiser
does platform independent optimisations. The back-end will generate the 
executable code. This architecture can be seen in Figure~\ref{fig:SimpleComp}.

LLVM has its own intermediate representation. There are multiple
compilers that can generate LLVM IR. These compilers are utilising
the same optimisation and code generation libraries. This approach
significantly lowers the barriers writing a new industrial strength
optimising compiler and also helps the community to focus on
better tools by sharing the work on the middle end and the
back-end among the developers of different compilers.

\begin{figure*}
\centering
\includegraphics[width=160mm]{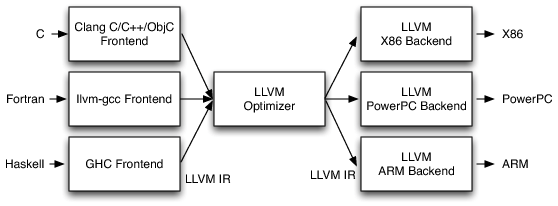}
\caption{Viewing a bug path in CodeChecker}
\label{fig:LLVMComp}
\end{figure*}

This architecture also improves the interoperability between
languages. Combining the LLVM IR from multiple front-ends might
enable optimisations that span across calls between different
languages. The architecture can be seen on Figure~\ref{fig:LLVMComp}.

The Clang compiler is also part of the LLVM Project.

The optimiser is doing fair amount of static analysis on the 
LLVM IR. Two of the major tasks of optimisations are the
following: establishing a cost model and approximate the
execution time of a piece of code and proving that a
transformation does not change the semantics of the code.
If a transformation is safe and it improves the performance
of the program according to the cost model the optimiser has,
LLVM will apply the transformation.

There are lots of different transformations within LLVM
and they are called passes. The order of passes has a
significant effect on the end result. It is also not
uncommon to run a pass multiple times or have a pass
whose sole purpose is to make code more
suitable for the subsequent passes.

\section{The Challenges of Static Analysis}

Due to the limitations of static analysis, in general, it is inevitable to either have 
false positives (incorrectly reporting an error) or false negatives (missing errors).
In software verification and mission critical
applications, it is more important to catch all possible errors. For a practical
bug finding tool, however, it is more important to have few false alarms, so
developers will take the results seriously. Usually, it is possible to reduce the
number of false negative results at the cost of increasing the number of false
positive ones and vice-versa. The classification of defects found
by an analysis tool can be seen in Table~\ref{fig:bes}.

\begin{figure}
\begin{center}
  \begin{tabular}{ | l | c | r | }
    \hline
                      & Bug exists     & Correct code     \\ \hline
    Error reported    & True Positive  & False Positive   \\ \hline
    No error reported & False Negative & True Negative    \\
    \hline
  \end{tabular}
\end{center}
\caption{Classification of the results of static analysis}
\label{fig:bes}
\end{figure}

The halting problem is the problem of determining, from a description of an
arbitrary program and an input, whether the program terminates or runs
forever. A general program to solve this problem for all program-input
pairs cannot exist. Since it is infeasible to automatically prove the correctness
of large-scale programs, analyses make unsound assumptions due to approximations.
The implications of these approximations are the false positive and false negative
reports. There are several techniques to deal with false positives. It is possible
to rank the results and make important violations appear first. Statistical methods
and other heuristics are also often used.

Fortunately, the approximations that the analyses are using work well most of the
time. Code containing complex data and control dependencies are rare, because
they are also difficult for programmers to understand, so they avoid them as
bad practice.

Let us investigate why it is impossible to solve the halting problem.
To prove this we need to show that there is no total computable function that
decides whether an arbitrary program $i$ halts on arbitrary input $x$:

\[ h(i, x) =
  \begin{cases}
    1       & \quad \text{if } i \text{ halts on input } x\\
    0       & \quad \text{otherwise}\\
  \end{cases}
\]

Where $i$ refers to the $i$th program in an enumeration of all programs.
We can show that every total computable function with two arguments differs
from $h$.

Given any total computable binary function, $g$ is a computable partial
function:

\[ g(i) =
  \begin{cases}
    0            & \quad \text{if } f(i,i) = 0\\
    undefined    & \quad \text{otherwise}\\
  \end{cases}
\]

Since $g$ is computable, there must be a program $e$, that computes $g$. We can
show that $h(e, e)$ has a different value than $f(e, e)$. It follows from the
definition of $g$ that exactly one of the following two cases must hold:

\begin{itemize}
  \item $f(e, e)=0$ so $g(e)=0$. In this case $h(e, e)=1$.
  \item $f(e, e) \neq 0$ so $g(e)$ is undefined. In this case $h(e, e)=0$.
\end{itemize}

In either case $f$ must be different from $h$. Because the choice of $f$ was
arbitrary, $h$ cannot be computable.

Rice's theorem from 1953 can be informally paraphrased as all interesting
questions about program behaviour are undecidable. This can be easily
proved for any special case. Assume the existence of an analyser that can check
whether a variable has a constant value during the execution of the program.
We could use this analyser to decide on the halting problem by using an input
program such as Figure~\ref{fig:const_check} where $tm$ returns $true$ if and only if
the $j$th program in the enumeration terminates on empty input.

\begin{figure}
\begin{lstlisting}
int x = 5;
if (tm(j))
  x = 6;
\end{lstlisting}
\caption{Input for const-ness check}
\label{fig:const_check}
\end{figure}

This can be a very discouraging result, but most of the time the real focus of
static analysis is not to prove such properties but rather to solve practical
problems. This means that such analysis can often produce approximate results.

\section{Challenges Implied by Languages}

The syntax and the type system of a language can have a significant effect on the
compilation time and the complexity of language-related tools. The compilation
model has similar impacts. Programs written in dynamically typed languages tend
to convey much less information statically about the properties of the program,
making the static analysis much more challenging. In order to overcome this 
obstacle, it is common to augment static analysis with information that was
gathered during an execution of the program.

The efficiency of static analysis tools depends on how easy it is to parse
a language. The main challenge in parsing is the ambiguities. In order to solve
these ambiguities, the parser needs to look ahead and read additional tokens.
The minimum number of times a token needs to be inspected in the worst case
depends on the grammar of the language. Designing the language grammar with
these problems in mind can make the life of compiler and tool authors easier.

\begin{figure}
\begin{lstlisting}
S (x);
T * y;
func((R) * z);
T T;
\end{lstlisting}
\caption{Ambiguous C grammar}
\label{fig:ctxt_dep_gram}
\end{figure}

The grammar of C and C++ is ambiguous without sufficient type information.
There are some examples of ambiguity in Figure~\ref{fig:ctxt_dep_gram}.
If $S$ is a type, then the first line is the definition of a variable, otherwise, if
$S$ is a function, then that expression is a function call. The second line can
be interpreted as a definition of a pointer or a multiplication depending on the
type context. The third line can be a multiplication or a dereference followed
by a C type cast. The last line shows that it is possible to have the same name
as both a variable name and a type name in the same scope.
To get the type information the headers included in a file need to be parsed
before the file can be parsed. 

One of the factors that makes static analysis harder is the number of dynamic
bindings. In case a virtual method is invoked, the analyser cannot know which
method will be executed unless the allocation of the receiver object is known.

The type system also has an important role in the analysis. The
purpose of the type system is to prove certain properties of the program. The
process of type checking is also static analysis. But in case one develops a 
separate analysis it is often assumed that the analysed program is correctly
typed. This assumption provides the analysis with much useful information.
The more expressive the type system is the more information is likely to be
available at compile time. Note that, type checking has the same limitations as
any other kind of static analysis. There are strongly normalising type systems
that guarantee the termination of all well-typed programs. Unfortunately,
either the expressivity of those systems is very limited or the type checking
of an arbitrary program is undecidable.

\section{Dynamic Analysis}

In dynamic analysis special code is generated and sometimes a run-time library
or virtual execution environment is also used to analyse the run-time behaviour
of the program.

Dynamic analysis has a run-time cost but it can also be more precise. During 
execution much more information is available, it is possible to achieve fewer
false positives. On the other hand only executed code can be analysed. In case
a bug is not triggered or a piece of code is not covered we will miss the 
problem. Static analysis has the possibility to discover errors in uncovered
code or find unexpected errors that we did not write a test for.

Also, there are some properties of the code that can not be tested or analysed
dynamically. These properties include naming conventions, code formatting,
misusing some language features, code smells, and portability issues.\

LLVM has a number of dynamic analysis tools, the sanitisers. These can find
memory addressing issues, memory leaks, race conditions, and other sources
of undefined behaviour like division by zero.
There are other popular tools outside of the LLVM project such as Valgrind.

We do not cover the details of dynamic analysis in this class but it is a useful
tool to have at your disposal. Static and dynamic analysis are two different
tools for slightly different purposes. It is better to use both techniques to
get the best of both worlds.

\section{Methods of Static Analysis}

There are several methods to analyse software statically. The easiest way is to
transform the code into a canonical form, tokenise it, and then use regular
expressions or other pattern matching method on that token stream. This method
is efficient and it is used in CppCheck. Unfortunately, it has a lot of
weaknesses as well, for example, the type information cannot be utilised
easily. It is also hard to match for implicit constructs in the language, for
example implicit casts.

A more advanced approach is to use the AST of the 
source code and match patterns on the tree representation. The tree can be 
handled more easily than a token stream. In case this AST is typed, type
information and implicit constructs are available as well. One of the 
disadvantages of this approach is that it is hard to do any control flow-dependent
matching on the AST. For example, it is hard to reason about the
values of the variables in this representation. For some class of analysis this
approach is sufficient. Clang Tidy is using this approach.

In a flow-sensitive analysis the analyser builds the CFG.
A flow-sensitive analysis is a polynomial analysis
that traverses the CFG and analyses each node during the traversal. Each time
two branches fold into one, the information gathered on those branches are
merged. This analysis is powerful enough to catch issues like non-trivial cases
of division by zero.

\begin{figure}
\begin{lstlisting}
int i;

if (a) {
  i = 0;
} else {
  i = 1;
}
...
if (!a) {
  j = 5/i;
} else {
  j = 3/i;
}
j += 2/i;
\end{lstlisting}
\caption{Example for division by zero}
\label{fig:div_zero}
\end{figure}

Unfortunately, flow-sensitive analysis is not precise enough for some purposes.
Consider an analysis that checks for division by zero errors. It would
track the values of the variables and store which ones are zero. What should
such analysis do, when two branches fold together? One option would be to mark
the value of a variable zero when it was zero on any of the branches. If you
look at Figure~\ref{fig:div_zero}, there will be a false positive case when
$5$ is divided by $i$, since this branch is only taken when $i$ is not zero.
This decision makes this analysis an over-approximation. The other option would
be to mark the variable as zero when it is marked as zero on both of the
branches. This analysis marks none of the variables as zero in
Figure~\ref{fig:div_zero}.
There is a false negative in the last line of the code snippet in this case,
this makes the analysis an under-approximation.

Obviously, there is room for improvement which is called path-sensitive
analysis. In path-sensitive analysis, each possible execution path of the program
is considered. This method this method can achieve a great precision. The analysis can exclude
some impossible branches and it can track the possible values of the
variables more precisely. There is a cost to this precision in execution
time, as this kind of analysis analysis has exponential complexity in the number of branches. 
Fortunately, the run-time of such analysis turns out to be manageable in practice.
There are existing tools that use this method on industrial scale codebase.
In case the analyser tracks the possible values of variables as symbols and
does symbolic computation with them, we are talking about symbolic execution.
To track the possible values of those symbols a
constraint manager and a constraint solver are used. One of the main sources of those 
constraints are conditions on the branches of the CFG.

Symbolic execution is a way of performing abstract interpretation. Abstract
interpretation is a way of interpreting programs, but instead of using the
concrete semantics (the most precise interpretation) of the program we use
an abstract semantics. The reason of defining abstract semantics is to make the
problem more tractable.

Software semantics are not always determined by the program code. It can
also depend on compilation parameters. For this reason, it is crucial to know
those parameters during abstract interpretation. Those parameters can be macro
definitions, the language standard choice, platform etc. This makes determining
such parameters part of the static analysis in most of the cases. Most of the
time in practice this is achieved by the logging of the arguments passed to the
compiler during a compilation.

Note that there are several formal methods (a lot of them are based on abstract
interpretation) that are used to verify the correctness of software, but they are
out of the scope of the class. It focuses on practical tools with large-scale
industrial applications.

Once we have flow-sensitive or path-sensitive analysis it is interesting to 
consider what is the biggest scope that the analyser can reason about using
these approaches.

\section{Clang Static Analyzer}
\label{csa}

The Clang SA uses symbolic execution to analyse C, C++, and Objective-C code.
During symbolic execution, the source code is interpreted, each unknown value
is represented with a symbol, the calculations are carried out symbolically.
During interpretation, the SA attempts to enumerate all possible execution paths.
We call such analyses path-sensitive.
To represent the analysis SA uses a data structure called \emph{exploded graph}
\cite{Reps95}. Each vertex of this graph is a \texttt{(symbolic state, program point)} pair.
A \emph{symbolic state} can correspond to a set of real program states. The \emph{program point}
determines the current location in the program, it is similar to an instruction
pointer. The graph edges are transitions between the vertices.
Memory is represented using a hierarchy of memory regions \cite{Xu2010}.
The analyser is building the graph on demand during the analysis using a depth-first
method with a path-sensitive walk over the control flow graph. The vertices of the
control flow graph are basic blocks which are code segments that are always executed
sequentially. The edges are possible jumps between these blocks.

The symbolic state consists of three components:
\begin{itemize}
\item \textbf{Environment} -- a mapping from source code expressions to symbolic expressions.
\item \textbf{Store} -- a mapping from memory locations to symbolic expressions.
\item \textbf{Generic Data Map (GDM)} -- checks can store domain specific information here.
\end{itemize}

During the execution of a path, the analyser collects constraints on the
symbolic expressions. There is a built-in constraint solver that represents
these constraints using alternation of ranges. This solver can reason about
pointers and integers. It is also possible to use
\emph{Z3} \cite{deMoura2008} as an external solver. These constraints are used to
skip the analysis of infeasible paths. The constraint solver can also be
utilised by the checks to query certain information about the program states.
This functionality can be used, for instance, to detect out of bounds
errors.

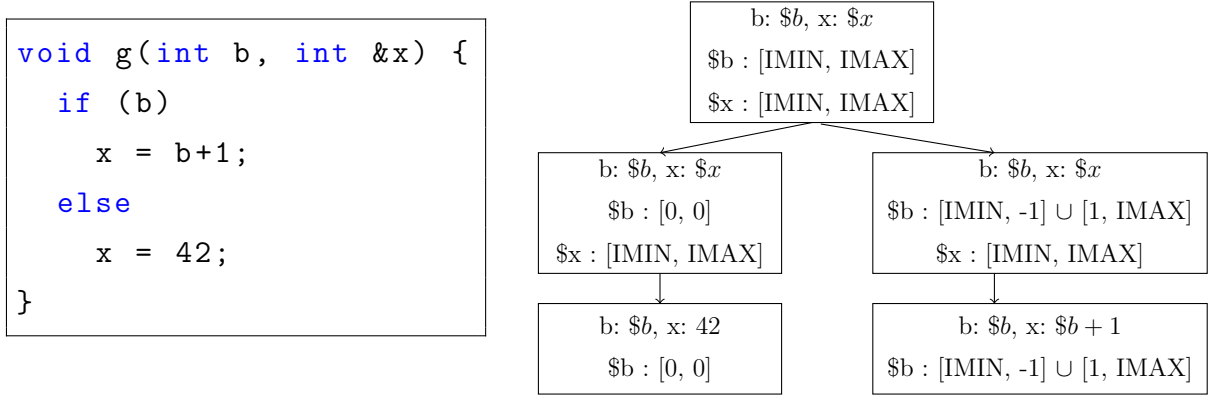
\begin{figure*}
\noindent\begin{minipage}{.38\textwidth}
\begin{lstlisting}[frame=tlrb]{Name}
void g(int b, int &x) {
  if (b)
    x = b+1;
  else
    x = 42;
}
\end{lstlisting}
\end{minipage}\hfill
\begin{minipage}{.57\textwidth}
\begin{tikzpicture}[scale=0.8, transform shape]
\draw  (-2.5,3) rectangle (1.5,1) node[pos=.5, rectangle split,rectangle split parts=3] {b: $\$b$, x: $\$x$\nodepart{second} \$b : [IMIN, IMAX]\nodepart{third} \$x : [IMIN, IMAX]};
\draw  (-5,0.5) rectangle (-1,-1.5) node[pos=.5, rectangle split,rectangle split parts=3] {b: $\$b$, x: $\$x$\nodepart{second} \$b : [0, 0]\nodepart{third} \$x : [IMIN, IMAX]};
\draw  (-5,-2) rectangle (-1,-3.5) node[pos=.5, rectangle split,rectangle split parts=2] {b: $\$b$, x: $42$\nodepart{second} \$b : [0, 0]};
\draw  (0.5,0.5) rectangle (6,-1.5) node[pos=.5, rectangle split,rectangle split parts=3] {b: $\$b$, x: $\$x$\nodepart{second} \$b : [IMIN, -1]$\ \cup\ $[1, IMAX]\nodepart{third} \$x : [IMIN, IMAX]};
\draw  (0.5,-2) rectangle (6,-3.5) node[pos=.5, rectangle split,rectangle split parts=2] {b: $\$b$, x: $\$b+1$\nodepart{second} \$b : [IMIN, -1]$\ \cup\ $[1, IMAX]};
\draw [->](-0.5,1) node (v1) {} -- (-3,0.5);
\draw [->](v1) -- (2.5,0.5);
\draw [->](-3,-1.5) -- (-3,-2);
\draw [->](2.5,-1.5) -- (2.5,-2);
\end{tikzpicture}
\end{minipage}
\caption{Symbolic execution and exploded graph}
\label{fig:exploded}
\end{figure*}

An example analysis can be seen along with its exploded graph in Figure~\ref{fig:exploded}.
Function \texttt{g} contains two execution paths. Since the values of \texttt{b}
and \texttt{x} are initially unknown, these values are represented with the
corresponding symbols \texttt{\$b} and \texttt{\$x}. These symbols can have
arbitrary values. As the analysis continues on one of the execution paths the
value of \texttt{b} is known to be zero and later on this path we discover that
the value of x is the constant \texttt{42}. The symbol \texttt{\$x} is no longer
needed on this path. On the second path, the value of \texttt{b} can be anything
but zero. On this path, we also discover that the value of \texttt{x} is greater by one
than the original value of \texttt{b}. The symbol \texttt{\$x} is no longer needed
on any of the paths, it can be garbage collected.

If the analyser finds a critical issue such as division by zero it reports
the issue to the user and also stops the analysis on that execution path. The
reason is that there is no meaningful way to simulate the program execution
after a critical error happened. The finding is presented along with the
execution path to make it easier to understand how a program execution can
lead to the error.

The SA also supports interprocedural analysis in a context-sensitive way.
When there is a call to a function with a known body, the analyser can
continue the analysis inside the callee preserving all the information
known at the call site. This is called \emph{inline analysis}.

The analyser cannot know where to start the analysis. The \texttt{main} function might
not be available or in case of libraries, it might not exist at all. SA usually picks
a function and starts to interpret it without calling context. Such function is called
a \emph{top level function}. If a function body is not available it will evaluate it
conservatively by losing precision otherwise it will do inline analysis. After
the analysis is done, the analyser will pick a new top level function which was
not visited (inlined) during the analysis of earlier top level functions.

\begin{figure}
\begin{lstlisting}
void h(int x) { /* ... */ }
void f(int &x);

void g(int x) {
	if (x > 0) {
		h(x);
		f(x);
		h(x);
	}
}
\end{lstlisting}
\caption{Loss of precision}
\label{fig:tu_prob}
\end{figure}

Let us look at Figure~\ref{fig:tu_prob}. The definition of $f$ is in a separate translation
unit, the definition of $h$, however, is available. When the analyser evaluates
$g$ it creates a symbol to represent the unknown value of $x$.
It collects the constraints on this symbol as alternation of intervals.
It knows that at the point at which $h$ is called for the first time the value of $x$ is 
positive due to the constraint in the \texttt{if} statement. The value of $x$ after the
call to $f$ is unknown, because $f$ might modify it. Consequently, we have
less amount of information about the program state at the second call to $h$.
This loss of precision can be quite severe. In case an element of an array or a member
of a struct is passed to $f$, the whole array or struct is marked as unknown,
because $f$ might use pointer arithmetic to modify elements other than the one
that was passed to it.

This loss of precision can cause both false positive and false negative results.
For example the analyser might not catch a division by zero error, because the
information about the variable which has zero value on that path might be erased
by a call to an unknown function. False positives are the result of analysing
more impossible branches due to a lack of information.

It is intractable to enumerate all the execution paths of non-trivial programs
as the number of paths is exponential in the branching factor. Loops make
this problem even worse as potentially every iteration might involve multiple
branches. To circumvent
this problem the analyser employs heuristics to limit the size of the
generated exploded graph and keep the run-time of the analysis within reasonable
limits. Here are some of the applied approaches:
\begin{itemize}
\item After some call depth, functions will not be inlined but evaluated 
  conservatively, even when the body is available. Small enough functions will always
  be inlined regardless of this limit.
\item Functions with too large control flow graphs will be inlined only a limited number of times.
\item If a basic block is visited more than some threshold on a path the analysis will terminate
  on that path. This is particularly useful for loops.
\item If the number of generated exploded graph vertices exceeds a threshold the
  analysis of the top level function will be terminated.
\end{itemize}

We call the set of thresholds together analysis budget.
The heuristics above were designed to cover a reasonable amount of execution
paths in a reasonable time while maintaining a good coverage of the code.
If we want to extend the scope of the analysis, it is important to ensure
that we do not suffer significant coverage loss due to running out of
analysis budget before finding the interesting issues.

\section{CodeChecker}

CodeChecker is a static analysis infrastructure built on the LLVM/Clang Static
Analyzer tool-chain, replacing scan-build in a Linux or macOS (OS X) development environment.
Its main features include:
\begin{itemize}
\item Support for multiple analysers, currently Clang Static Analyzer and Clang-Tidy
\item Store results of multiple large-scale analysis runs efficiently, either in a PostgreSQL or SQLite database
\item Web application for viewing discovered code defects with a streamlined, easy experience
\item Filterable (defect checker name, severity, source paths, ...) and comparable (calculates difference between two analyses of the project, showing which bugs have been fixed and which are newly introduced) result viewing
\item Subsequent analysis runs only check and update results for modified files without analysing the entire project (depends on build tool-chain support!)
\item See the list of bugs that have been introduced since your last analyser execution
\item Suppression of known false positive results, either using a configuration file or via annotation in source code, along with the exclusion of entire source paths from analysis
\item Results can be shared with fellow developers, the comments and review system aids the communication of code defects
\item Can show analysis results on standard output
\item Easily implementable Thrift-based server-client communication used for the storage and query of discovered defects
\item Support for multiple bug visualisation front-ends, such as the a application, a command-line tool, and an Eclipse plugin
\end{itemize}

\subsection{Quick Tutorial}
First add CodeChecker to your \texttt{PATH} and activate the python environment.

Analyse your project with the check command:

\begin{lstlisting}
CodeChecker check -b "make clean && make" -o ~/results
\end{lstlisting}

\texttt{check} will print an overview of the issues found in your project by the analysers.

Start a CodeChecker web and storage server in another terminal or as a background process. By default it will listen on \texttt{localhost:8001}.

The SQLite database containing the reports will be placed in your workspace directory (\texttt{~/.codechecker} by default), which can be provided via the \texttt{-w} flag.

\begin{lstlisting}
CodeChecker server
\end{lstlisting}

Store your analysis reports onto the server to be able to use the Web Viewer.

\begin{lstlisting}
CodeChecker store ~/results -n my-project
\end{lstlisting}

Open the CodeChecker Web Viewer in your browser, and you should be greeted with a web application showing you the analysis results.

For more information see the GitHub page.

\begin{figure*}
\centering
\includegraphics[width=160mm]{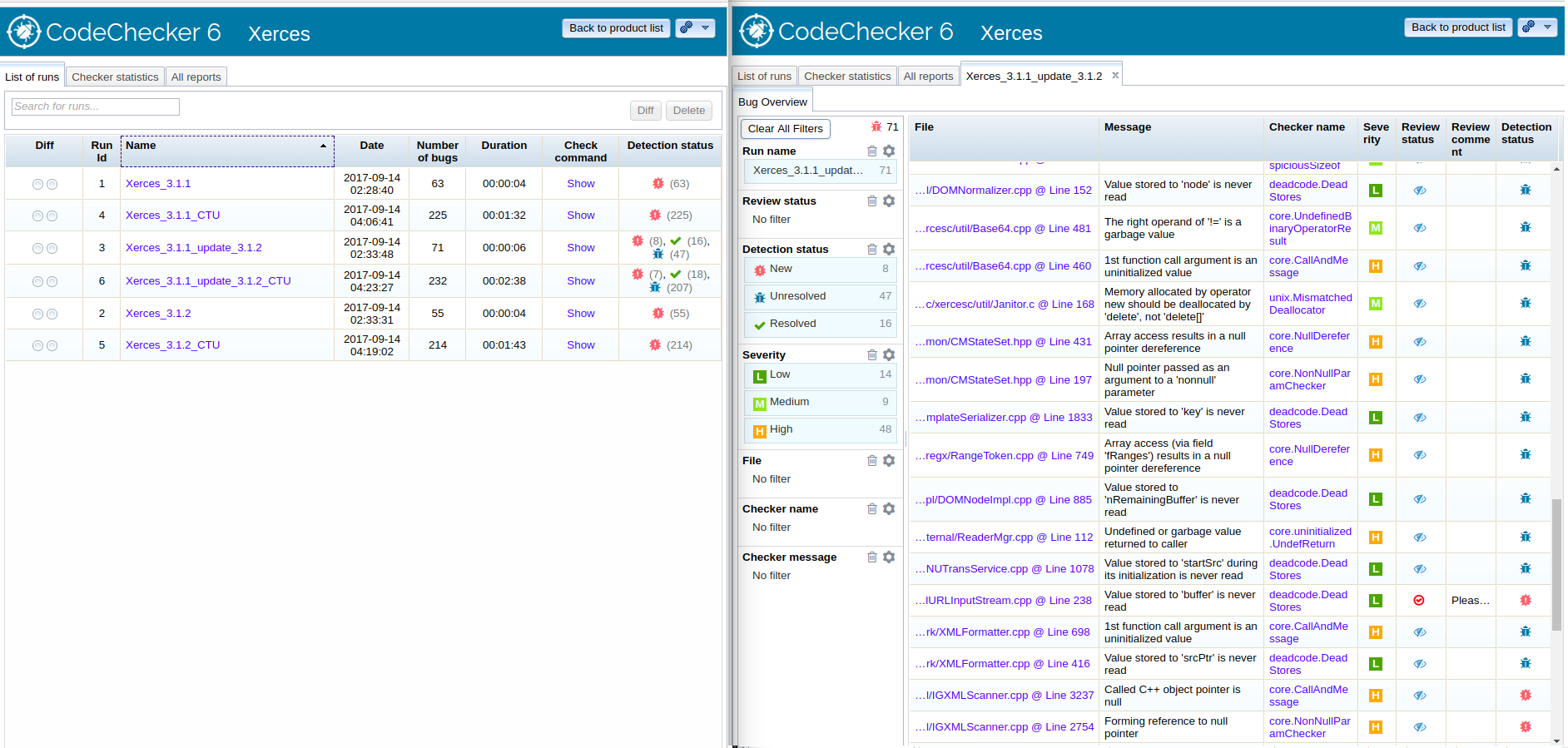}
\caption{Bug list view in CodeChecker}
\label{fig:BugList}
\end{figure*}

\begin{figure*}
\centering
\includegraphics[width=160mm]{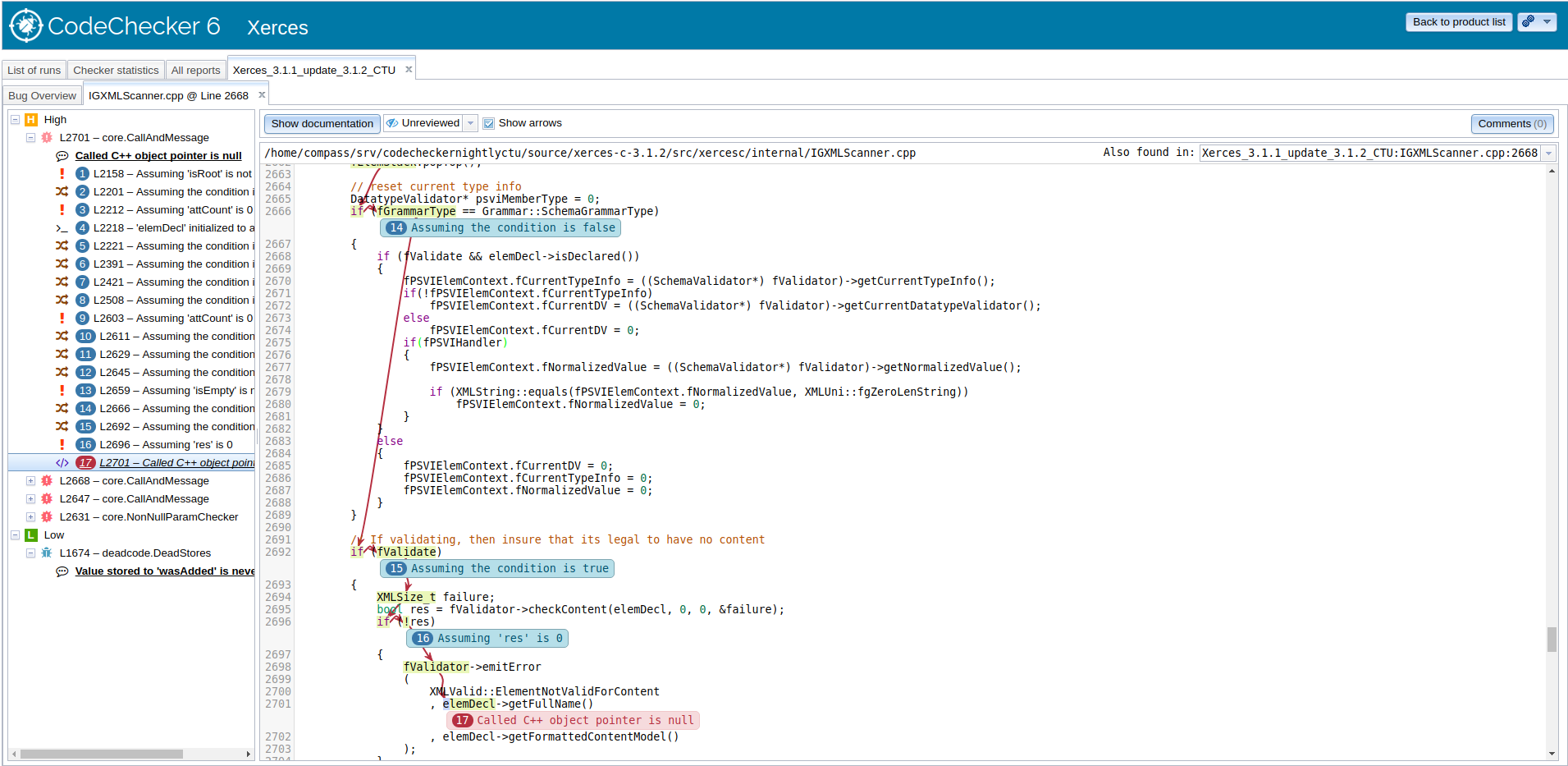}
\caption{Viewing a bug path in CodeChecker}
\label{fig:BugPath}
\end{figure*}

\pagebreak

\bibliography{3cows-bibliography}

\end{document}